\documentclass{article}[12pt]

\usepackage{amssymb}
\usepackage{graphicx}

\def\eq{\begin{equation}}
\def\en{\end{equation}}
\def\eqa{\begin{eqnarray}}
\def\ena{\end{eqnarray}}
\newcommand{\me}{\mathrm{e}}
\newcommand{\dif}{d}
\def\partialby#1{\frac{\partial\hfill}{\partial#1}}
\def\Hzero{H}
\def\zboundary{z}
\def\TL{T_{L}}
\def\TR{T_{R}}
\def\kB{k}
\def\bra#1{\langle #1 |}
\def\ket#1{| #1\rangle}
\def\expval#1{\langle \, #1 \,\rangle}
\def\expvalequil#1{\expval{#1}_{\mathit{eq}}}
\def\expvalc#1{\expval{#1}_{\mathit{c}}}
\def\comm#1#2{{[}#1 ,\; #2{]}}
\def\Real{{\mathbf{Re}}}
\def\Imag{{\mathbf{Im}}}

\def\chiren{\chi^{\mathit{ren}}}
\def\geucl{g}
\def\calB{{\cal B}}

\def\thefootnote{\fnsymbol{footnote}}
\begin{document}
\begin{titlepage}
\today          \hfill
\begin{center}
\hfill hep-th/0512023  \\

\vskip .5in
\renewcommand{\thefootnote}{\fnsymbol{footnote}}
{\Large \bf
Infrared properties of boundaries in 1-d quantum systems
}

\vskip .50in

\vskip .5in
{\large Daniel Friedan}\footnote{email address: friedan@physics.rutgers.edu}
and
{\large Anatoly Konechny}\footnote{email address: anatolyk@physics.rutgers.edu}

\vskip 0.5cm
{\large \em Department of Physics and Astronomy,\\
Rutgers, The State University of New Jersey,\\
Piscataway, New Jersey 08854-8019 U.S.A.} \\

\end{center}

\vskip .5in

\begin{abstract} \large
We present some partial results on
the general infrared behavior of
bulk-critical 1-d quantum systems with boundary.
We investigate whether
the boundary entropy, $s(T)$,
is always bounded below as the temperature $T$ decreases towards $0$,
and whether the boundary always becomes
critical in the IR limit.
We show that
failure of these properties
is equivalent to certain seemingly
pathological behaviors far from the
boundary.
One of our approaches uses real time methods, in which
locality at the boundary is expressed by analyticity in the
frequency.
As a preliminary,
we use real time methods to
prove again that the boundary beta-function
is the gradient of the boundary entropy,
which implies that $s(T)$ decreases with T.
The metric on the space of boundary couplings is interpreted as the
renormalized susceptibility matrix of the boundary,
made finite by a natural subtraction.
\end{abstract}
\end{titlepage}
\large

\newpage
\renewcommand{\thepage}{\arabic{page}}
\setcounter{page}{1}
\setcounter{footnote}{0}
\renewcommand{\thefootnote}{\arabic{footnote}}
\large
\section{Introduction}
In this paper we study the infrared behavior of
bulk-critical one-dimensional quantum systems with boundary.
These are 1-d quantum systems whose bulk couplings are at a
critical point, but whose boundary couplings are
not necessarily critical.  We would like to show that the
boundary couplings are always driven to a renormalization
group fixed point in the far infrared, which is to say that
the boundary always becomes critical in the infrared limit.
We would also like to show that the boundary entropy cannot
decrease without limit, but must approach some lower bound
as the temperature decreases towards zero.  Alternatively,
we would like to understand what kind of quantum boundary
does \emph{not} go to an IR fixed point, or \emph{does}
release an unlimited amount of entropy as its temperature
goes to zero.  We record here some partial results which
might be useful as steps towards these goals.

The boundary entropy, $s$, is the difference between the
total entropy and the bulk entropy (which is
proportional to the length of the system).  For critical
boundaries, the number $g=\exp (s)$ is the universal non-integer ground state
degeneracy of Affleck and Ludwig\cite{AL1}.
In \cite{FK}, we proved a gradient formula
\eq
\frac{\partial s}{\partial \lambda^{a}}
= - g_{ab}(\lambda) \beta^{b}(\lambda)
\label{eq:gradientformula}
\en
which expresses the boundary beta-function, $\beta^{b}$, as
the gradient of the boundary entropy, $s$, with respect to a
certain metric, $g_{ab}$, on the space of all the marginal
and relevant boundary couplings.  The $\lambda^{a}$
are the boundary coupling constants.  The
boundary entropy depends on the temperature and the
boundary couplings, and satisfies the renormalization group
equation
\eq
\left (T\partialby{T} + \beta^{a}\partialby{\lambda^{a}}
\right ) s = 0
\,,
\en
so the boundary gradient formula implies that
\eq
T\frac{\partial s}{\partial T} =
\beta^{a} g_{ab} \beta^{b} \ge 0\,.
\en
Thus $s(T)$ always decreases with decreasing
temperature, which is to say that the boundary entropy always
decreases under
the renormalization group.
The boundary is
critical, $\beta^{a}(\lambda)=0$, if and only if
the boundary entropy is
stationary in the temperature,
$ds/dT =0$.
The boundary entropy can decrease below zero because
the third law of thermodynamics does not apply.
The boundary is not an isolated system.

We would like to understand the properties of the boundary in
the far infrared.  For bulk 1-d quantum systems, without
boundary,
the $c$-theorem\cite{Zamolodchikov} gives
considerable control over the infrared behavior.
The $c$-theorem states that a certain
function of the bulk couplings
decreases under the renormalization group,
is stationary if and only if the bulk beta-function
vanishes, and cannot become negative.
This is almost enough to show
that the bulk system must flow to a fixed
point in the infrared.
We point out below an additional assumption that is needed.

The generic bulk system has a mass gap, so it
flows in the infrared to the trivial $c=0$ fixed point,
where no excitations remain.  There does not seem to be
an analogously trivial
boundary system.  A boundary
that flowed to $s=-\infty$, $g=0$ might provide a
candidate, but no such system is known.
In every known example,
the infrared limit
is a non-trivial boundary fixed
point and
the boundary entropy decreases to a
finite lower limit.  Nontrivial boundary
excitations always remain.  It can be conjectured that
the boundary entropy is necessarily
bounded below throughout a RG flow, and
that the flow necessarily ends at an IR fixed point, unless some
pathologies develop.  We would like to understand what technical
assumptions are needed to prove these conjectures,
and what physical principles they express.

The boundary gradient formula succeeds in excluding some
exotic forms of renormalization group behavior.  For
example, limit cycles within the space of boundary couplings
are impossible.  But the gradient formula by itself does not
guarantee that the system flows to an infrared fixed point.
The boundary entropy might decrease without bound, with
$\beta^{a}(\lambda)$ never approaching zero.  This
possibility could be excluded if we could show that the
boundary entropy is bounded below (for a given bulk critical
system).
We are at least able to show, under certain assumptions, that
$\beta^{a}
g_{ab}\beta^{b}\rightarrow 0$ in the infrared limit (see
section~\ref{sect:estimate}).  This
is analogous to what the $c$-theorem provides in the
bulk.  It does not establish that $\beta^{a}$
vanishes in the infrared limit, just as the analogous bulk result
does not, but this is a step in the right direction.

A lower bound on the boundary entropy would also be of
interest because it would imply that only a bounded amount
of information can be added to a given boundary or junction
within a near-critical quantum circuit.  Such circuits have
been argued to be
the ideal physical systems for asymptotically large-scale
quantum computers\cite{FriedanFlowA}.  A lower bound on the boundary
entropy would be a very general constraint on the design of
such quantum computers.

There are a number of examples of a lower bound on $g=\exp
(s)$ for boundary conformal field theories corresponding to a
given bulk conformal field theory.  For the compact $U(1)$
Gaussian model with target radius $R$, normalized so that
$R=1$ is the self-dual radius, the lowest value of $s$
corresponds to the Dirichlet boundary condition,
$s_{D}=-\frac14 \ln 2 -\frac12 \ln R$, when $R\ge 1$, and to
the Neumann boundary condition, $s_{N}=-\frac14 \ln 2
+\frac12 \ln R$, when $R\le 1$, so the lower bound on $s$ is
$-\frac14 \ln 2 -\frac12 |\ln R|$.  Clearly, there is no
universal lower bound, independent of the bulk conformal
field theory.

Another set of examples are the conformal boundary
conditions given by the Cardy boundary states in rational
conformal field theories\cite{Cardy}.  Each Cardy boundary
state is labelled by a primary field $i$.  We point out in
the appendix that the Cardy state with the smallest value of
$s$ is the one associated with the identity operator, $i=0$,
so the lower bound on $s$ is $s_{0}= \frac12 \ln S_{00}$
where $S_{00}$ is the corresponding entry of the modular
$S$-matrix.  In the case of the unitary $c<1$ conformal
field theories, the Cardy states are all the possible
conformal boundary conditions. For the unitary minimal models
with  central charge
$$
c_{m} = 1 - \frac{6}{m(m+1)} \, , \enspace m = 2,3,\cdots
$$
the lower bound is
$$
s_{0}(m) = \frac{1}{4}\ln \left [ \frac{8}{m(m+1)}
\sin^{2}\left ( \frac\pi{m}\right ) \sin^{2}\left (\frac\pi{m+1}\right ) \right ] \, .
$$
In these examples, one can observe the
crucial role of locality in putting a lower bound on $s$.
It is the imposition of the Cardy constraint, which is a
form of the locality condition, that ensures a nonzero
overlap $g=\langle B| 0 \rangle$ between the boundary state
and the conformal vacuum.

In this paper, we start by arguing that any critical boundary system
must have $g>0$, or else the system would not have a sensible
thermodynamic limit.  We then argue that, for non-critical boundaries,
the boundary contribution, $\theta(\tau)$, to the trace of the
energy-momentum tensor goes to a multiple of the identity operator in
the far infrared.  We work directly at $T=0$.  Specifically, we show that its connected
two-point function in Euclidian time satisfies
\eq
\lim_{\tau\rightarrow\infty} \tau^{2}
\expvalc{\theta(\tau)\,\theta(0)} = 0
\:.
\label{thetatwopoint}
\en
We need to assume that, far from the boundary,
the bulk conformal invariance is
restored in a strong sense.
The canonical scaling
dimension of $\theta(\tau)$ is 1, so
equation (\ref{thetatwopoint})
comes close to implying that
$\theta(\tau)$ vanishes
up to a multiple of the identity operator,
which would imply that
the infrared limit is scale invariant.
To finish the argument, we need that the
correlation functions of the bulk operators satisfy a
cluster decomposition condition in the infrared limit.  This
is essentially the assumption that the infrared limit is a
well-defined boundary quantum field theory,
in which case the vanishing of the two-point function implies the
vanishing of the operator.
We do not know if our assumption is provable from general principles.
If this gap can be filled, then the
infrared limit at $T=0$ is a boundary quantum field theory with
$\theta(\tau)=\expval{\theta}\, {\bf 1}$, which is a boundary conformal field
theory.  Given the previous
argument that any boundary conformal field theory
has $s>-\infty$, the boundary entropy of the original
system would be bounded below.

An analogous gap exists in the argument that the infrared 
limit in the bulk is always a fixed point.  An assumption is 
also needed that the infrared limit is a well-defined 
quantum field theory, so that the vanishing of the two-point 
function of the trace of the energy-momentum tensor implies 
that the operator itself vanishes.  In the boundary case, 
the bulk operator algebra does not change under the 
renormalization group, so the situation might be better than 
in the bulk case.  This leaves a hope that our results can 
be strengthened.

Our second approach is to use real time methods at $T>0$.
As a preliminary step, we re-prove the boundary gradient formula
using real time methods, based on the spectral analysis of
the flow of entropy through the boundary\cite{FriedanFlowB}.  In this
version of the proof, the metric $g_{ab}$ is given a
physical interpretation.  It is the renormalized boundary
susceptibility matrix, made finite by a natural subtraction.
It can be measured experimentally.  We try to use the real
time formalism to show that $\dif s/\dif T
=\beta^{a}g_{ab}\beta^{b}/T$ is integrable with respect to $T$ at $T=0$.
This would imply a lower bound on $s$.
We only succeed in showing that $T\dif s/\dif T \to 0$
as
$T\rightarrow 0$,
which implies
that $\beta^{a}g_{ab}\beta^{b}\rightarrow 0$.
The condition of integrability at $T=0$
is reformulated as an estimate on the low temperature behavior of
a certain spectral function, an estimate that we do not know
how to prove.

\section{Notations and basic facts}
We will be using both real and Euclidean time descriptions
of a one-dimensional quantum system.  Space-time coordinates
are $(x,t)$, $x\ge 0$.  The boundary is at $x=0$.  The
Euclidean time is $\tau=it$.  The space-time metric is
$$
(\dif s)^{2} = -v^{2}(\dif t)^{2} + (\dif x)^{2}
= v^{2}(\dif \tau)^{2} + (\dif x)^{2}
$$
where $v$ is the velocity of ``light''.
The system is in equilibrium at temperature $T$.
The imaginary time correlation functions
are periodic in Euclidean time, with period $\beta = 1/T$
(in units with $\hbar = \kB = 1$).
The normalized equilibrium expectation values
are denoted by $\expvalequil{\mathcal{O}}$.
The connected two-point expectation values are
$
\expvalc{ \mathcal{O}_{1}\, \mathcal{O}_{2}}
=\expvalequil{ \mathcal{O}_{1}\, \mathcal{O}_{2}} -
\expvalequil{ \mathcal{O}_{1}}\expvalequil{\mathcal{O}_{2}}
$.
The energy-momentum tensor is $T^{\mu}_{\nu}(x,t)$.
Conservation of energy-momentum in the bulk is
expressed by
\eq
\partial_{\mu}T^{\mu}_{\nu}(x,t)=0 \qquad x>0 .
\en
The Hamiltonian is
\eq \label{H}
\Hzero = -\theta(t) + \int_{0}^{\infty} \dif x\,\,T^{t}_{t}(x,t)
\en
where $-\theta(t)$ is the boundary energy operator.
Energy conservation at the boundary is\footnote{In the present paper our conventions differ
from the ones in \cite{FK} in that the energy-momentum
components have canonical
dimensions,
instead of being dimensionless, as in \cite{FK}.
As a result, extra factors of the RG-scale $\mu$ are
present in various equations in \cite{FK}.}
\eq \label{boundaryc}
\partial_{t} \theta(t) = T^{x}_{t}(0,t) \, .
\en
The energy density $T^{t}_{t}(x,t)$
is the only component of the energy-momentum tensor that has a
boundary contribution.
See \cite{FK} for a more complete discussion
of the bulk+boundary energy-momentum tensor.

Bulk criticality is equivalent to local scale invariance in
the bulk:
\eq
\Theta(x,t) = T^{\mu}_{\mu}(x,t) =
T^{x}_{x}(x,t)+T^{t}_{t}(x,t) = 0 \qquad x>0 .
\en
The trace of the energy-momentum
tensor
is concentrated at the boundary,
\eq
\Theta(x,t) = \delta(x) \theta(t),
\en
and can be expanded in the boundary fields:
\eq
\theta(t) = \beta^{a} \phi_{a}(t)
\en
where the boundary operators $\phi_{a}(t)$ are the relevant
and marginal fields localized at the boundary.  The
coefficients $\beta^{a}$ comprise the boundary
beta-function.  The operators
$\phi_{a}(t)$ have
ultraviolet scaling dimensions all $\le 1$.  The boundary coupling
constants, $\lambda^{a}$, are related to the boundary fields,
$\phi_{a}(t)$, by
\eq
\frac{\partial Z }{\partial \lambda^{a}}
= \frac{\partial \zboundary }{\partial \lambda^{a}}
= \beta \expvalequil{\phi_{a}(0)}
\en
where $Z$ is the full partition function and
$z$ is the boundary partition function.
The definition of $z$ starts with a system of
finite length, $L$.  An arbitrary boundary condition
is imposed at $x=L$.
In the thermodynamic limit $L\to \infty$,
the full partition function, $Z_{L}$, factorizes into a bulk part and
a boundary part:
$$
\me^{-\pi c L /6\beta} Z_{L} \to zz'
$$
where $c$ is the bulk conformal central charge and the
constants $z$ and $z'$ are the boundary partition functions
of the boundaries at $x=0$ and $x=L$ respectively.  Only the
product $z z'$ is determined.  Unitarity of the quantum
system implies that all the products $z z'$ are real and positive,
for all pairs of boundary conditions.
We can take the boundary condition at $x=L$ to be the same
as the boundary condition at $x=0$ (strictly speaking, the
CPT transform of the boundary condition at $x=0$).  Then
$\me^{-\pi c L /6\beta}Z_{L} \to  |z|^{2}$.  Now we can
determine $z$ as the positive real square root of $|z|^{2}$.
This is consistent, because all the products $z z'$ are positive
real numbers.
We construct the system
on the infinite half-cylinder
with a single boundary at $x=0$
by taking the limit $L\to \infty$,
dividing by $z'$ to
eliminate dependence on the
boundary condition at $x=L$.
In terms of the bulk conformal field theory on the half-cylinder,
where the spatial coordinate is $v\tau$ and the Euclidean
time is $x/v$,
the boundary condition at $x=0$ is represented by
a boundary state $\bra{B}$,
while the ``boundary condition'' at $x=L$ is
represented by the bulk ground-state $\ket{0}$,
since all the excited states at $x=L$ are suppressed
exponentially in $L$.
The boundary partition function is the overlap
$z=\expval{B|0}$.
The logarithm of the full partition function then takes the form
\begin{equation}\label{zL}
\ln Z_{L} = \frac{c\pi}{6\beta}L + \ln z
\end{equation}
where $c\pi/6\beta$ is the universal ground state energy 
density of
the bulk conformal field theory.

The total entropy of the system
is $S_{L}= (1-\beta\partial/\partial\beta) Z_{L}$.
Removing the bulk contribution leaves the boundary entropy
\begin{equation}\label{s}
s = \left(1 - \beta\frac{\partial}{\partial \beta} \right) \ln z \, .
\end{equation}
The boundary entropy is a function of $\mu \beta$, where $\mu$ is the
renormalization scale.  It satisfies the renormalization
group equation
\eq
\left (-\mu\partialby{\mu} + \beta^{a}\partialby{\lambda^{a}}
\right ) s =
\left (-\beta\partialby{\beta} + \beta^{a}\partialby{\lambda^{a}}
\right ) s = 0
\:.
\en
For thermodynamic quantities,
the infrared limit $\mu\to \infty$ is equivalent to the zero
temperature limit $T\to 0$.
In this paper, we will avoid writing $\mu$ and $\mu\to\infty$.
Instead, when we study thermodynamic quantities, we
will take $T\to 0$, and use the second form of
the renormalization group equation for $s$.
When we study the quantum field theory at $T=0$, we will take
the IR limit by scaling all times and distances to infinity
in the correlation functions.

The boundary beta-function vanishes at a fixed point,
so $s$ is then a number, independent of temperature:
$s=\ln z =\ln g$.
where $g$ is the universal noninteger ground state
degeneracy of Affleck and Ludwig \cite{AL1}.  This is the
``ground state'' degeneracy because, being constant in $T$, it can
be evaluated at $T=0$.  For any finite $L$,
the energy spectrum
is discrete, so the ground state degeneracy is then
an integer.
The spectrum becomes continuous in the limit $L\to \infty$,
so the numerical factor $z=g$ can be an arbitrary
nonnegative number.  In particular, it is possible to have
$g<1$, $s<0$.

Affleck and Ludwig conjectured that the value of $g$ is 
larger at the ultraviolet fixed point of a renormalization 
group trajectory than at the infrared fixed 
point\cite{AL1,AL2}.  This \emph{$g$-theorem} was proved in 
\cite{FK} by proving the boundary gradient formula, equation 
(\ref{eq:gradientformula}).  The boundary gradient formula 
implies that the boundary entropy decreases with decreasing 
temperature, $\dif s/\dif T>0$, so the boundary entropy 
decreases along the renormalization group trajectory, so the 
value of $s=\ln g$ at the ultraviolet fixed point, at 
$T=\infty$, is greater than the value at the infrared fixed 
point, at $T=0$.  Ordinary entropy in statistical mechanics 
always decreases with temperature, but this is not obvious 
for the boundary entropy.  The total entropy $S_{L}$ of the 
system of length $L$ does goes down with temperature, 
trivially, but the large bulk contribution, $c\pi L/3\beta$, 
also decreases with temperature, so it is not obvious that 
the difference, the boundary entropy, decreases with 
temperature.

The metric in the gradient formula is
\eq
g_{ab} =
\int\limits_{0}^{\beta}d\tau\int\limits_{0}^{\beta}d\tau'
\,\langle \phi_{a}(\tau)\phi_{b}(\tau')\rangle_{c}
\left [
1 - \cos \left ( \frac{2 \pi (\tau-\tau')}{\beta}
\right )
\right ]
\en
so
\eq
\frac{\dif s}{\dif T}
=
\frac{1}{T}\int\limits_{0}^{\beta}d\tau\int\limits_{0}^{\beta}d\tau'
\,\langle \theta(\tau)\theta(\tau')\rangle_{c}
\left [
1 - \cos \left ( \frac{2 \pi (\tau-\tau')}{\beta}
\right )
\right ]
\, .
\en
Canonical ultraviolet behavior ensures that any 
non-universal contact terms in the two-point function have 
dimension at most $2$.  The factor $1 - \cos \left ( 2 \pi 
(\tau-\tau')/\beta \right )$ vanishes to second order at 
$\tau=\tau'$, so no contact terms contribute to the metric.  
The metric is thus finite and universal, assuming canonical 
ultraviolet behavior.  However, it is difficult to see a 
physical interpretation of the metric when it is written in 
this form, as an integral of a two-point function over 
Euclidean time.

Given bulk conformal invariance, the symmetric
energy-momentum tensor has
only two independent components:
\eqa
T^{x}_{t}(x,t)  = -v^{2} T^{t}_{x}(x,t) &=& \TR(x,t) - \TL(x,t) \nonumber \, ,\\
v T^{t}_{t}(x,t) = -v T^{x}_{x}(x,t) &=& \TR(x,t) + \TL(x,t)
\,.
\label{2}
\ena
The bulk conservation law implies that
$\TR(x,t)$ and $\TL(x,t)$ are chiral currents:
\eq
\TR(x,t) = \TR(x-vt)\, , \quad \TL(x,t) = \TL(x+vt) \,.
\label{chirality}
\en
They are related to the Virasoro operators in the ``closed string''
channel:
\eqa \label{Ln}
\TR(z)= T_{zz}(z) &=& -\frac{v^{2}}{2\pi} T(z)
= -\frac{2\pi}{\beta^{2}}
\sum_{n=-\infty}^{\infty} \me^{-2\pi n z/ v \beta} L_{n} \, , \nonumber \\
\TL(\bar z) = T_{\bar z\bar z}(\bar z)
&=& -\frac{v^{2}}{2\pi} \bar T(\bar z)
= -\frac{2\pi}{\beta^{2}}
\sum_{n=-\infty}^{\infty}
\me^{-2\pi n \bar z/v \beta} \bar L_{n}
\ena
where $z=x+iv\tau= x-vt$.
The coefficients are fixed by calculating the Hamiltonian in
the ``closed string'' channel,
where $v\tau$ is the spatial coordinate and $x/v$ the
Euclidean time:
$$
H_{\mathit{closed}}= \frac{2\pi}{\beta} (L_{0}+\bar L_{0})
=
\int_{0}^{\beta}\dif \tau \,\, v T_{x}^{x} \, .
$$
On the semi-infinite cylinder, the boundary condition at
$x=\infty$ is the bulk ground state, which satisfies
$L_{n}\ket{0}=\bar L_{n}\ket{0}=0$, $n\ge -1$.  This implies
that the bulk energy-momentum tensor, within correlation functions,
decreases at infinity as
\begin{equation}
T^{\mu}_{\nu}(x,\tau) \sim e^{-4\pi x/\beta}, \quad x\to \infty \, .
\end{equation}
Energy conservation at the boundary becomes
\eq
\partial_{t} \theta(t) = \TR(-vt) - \TL(vt) \, .
\en
Therefore
\eqa \label{th}
\theta(t) &=& \int_{-\infty}^{t} \dif t'\,\, \TR(-vt')
- \int_{-\infty}^{t} \dif t'\,\, \TL(vt') \nonumber \\
&=& \int_{t}^{\infty} \dif t'\,\, \TL(vt')
- \int_{t}^{\infty} \dif t'\,\, \TR(-vt') \, .
\ena
From (\ref{H}), (\ref{2}), (\ref{th}) we obtain
\eqa \label{H2}
\Hzero = \int_{-\infty}^{\infty} \dif t \,\, \TR(vt)
=\int_{-\infty}^{\infty} \dif t \,\, \TL(vt)
\ena


\section{Boundary entropy at a fixed point and locality}
\label{fp_locality}
We argue that all critical boundaries have
$g>0$.
Suppose otherwise.
Then there would be a conformal boundary condition given by a
boundary state $ \ket{B} $ such that
$g =\langle 0\ket{B} = 0 $.
In this case we should re-examine the $L\to \infty$
thermodynamic limit.  The boundary state
$\ket{B}$ is put at $x=0$.  At the
other boundary, at $x=L$, we put a boundary condition $\bra{B'}$.
We choose $\bra{B'}$ with the property that
$\bra{B'} 0\rangle >0$ so as to ensure
that a conformal vacuum remains when we take
the limit $L\to \infty$.  The
one point functions of bulk operators are then defined as
\begin{equation}
\langle \phi(x,t) \rangle = \lim_{L\to
\infty} \frac{\langle
B'|e^{-LH_{\mathit{closed}}}\phi(x,t)|B\rangle}{\langle
B'|e^{-L H_{\mathit{closed}}}|B\rangle} \, .
\end{equation}
The numerator in this fraction has goes as $\me^{L\pi c/6}$
for large $L$, while the denominator goes as $\me^{L(\pi c/6
- \Delta_{1})}$ where $\Delta_{1}$ is the lowest eigenvalue
occurring in the action of $H_{\mathit{closed}}$ on
$\ket{B}$.  If $\langle 0\ket{B} = 0$, then $\Delta_{1}>0$
and the above limit is infinite, which means that there is
no sensible thermodynamic limit.  Alternatively, we could
try defining a thermodynamic limit by putting the boundary
state $\ket{B}$ on both ends of the cylinder.  In the
limit $L\to\infty$, we would obtain
finite correlation functions on the infinite half-cylinder,
but these correlation functions would generically grow
exponentially with separation and thus violate cluster
decomposition in the $x$-direction.
So $g>0$ for any sensible boundary conformal field theory.

\section{At $T=0$, $\lim_{\tau \to
\infty}\tau^{2}\expvalc{\theta(\tau)\theta(0)} = 0$}

Next, we try to argue that every boundary system flows to an
infrared fixed point: a scale invariant, conformally
invariant boundary field theory.  Then, by the argument
above, the boundary entropy would necessarily be bounded
below, because the infrared fixed point would have $g>0$.

We work directly at $T=0$.  The Euclidean space-time is
the half plane, $x\ge 0$, $-\infty<\tau<\infty$.
We argue that
\begin{equation} \label{lim}
\lim_{|\tau-\tau'|\to \infty} |\tau-\tau'|^{2}\langle
\theta(\tau) \theta(\tau')\rangle_{c} = 0 \, .
\end{equation}
Here $\langle \theta(\tau)\theta(\tau')\rangle_{c}$ is
the zero temperature connected correlator evaluated on
the boundary of the infinite half-plane.
The factor
$|\tau-\tau'|^{2}$ accounts for the canonical scaling
dimension of $\theta(\tau)$.

If we can assume that the infrared limit is a boundary
quantum field theory, then we can conclude from
equation (\ref{lim}) that
$\theta(t)$ is a multiple of the identity in that limiting
theory, so the infrared limit is a conformally invariant
boundary quantum field theory, a fixed point of the
renormalization group.

The last assumption, that the infrared limit is a quantum
field theory, is also implicitly present when the $c$-theorem
is used to show that every bulk quantum field theory goes to
a fixed point (perhaps trivial) in the infrared.  The
$c$-theorem\cite{Zamolodchikov} implies that the trace,
$\Theta=T^{\mu}_{\mu}$, of the bulk energy-momentum tensor
has a vanishing connected 2-point function in the infrared
limit.  This in turn implies that all correlation functions
of the limiting theory are conformally invariant.  The
implicit assumption is that those correlation functions
exist in the infrared limit.  In the boundary case the
situation might be more favorable, because the bulk operator
algebra stays fixed (is not flowing).  This leaves a hope
that our results can be strengthened.

Our argument is based on the principle that the system
should become conformally invariant far from the boundary.
Consider the quantization in which
$\tau$ is the spatial
coordinate and $x$ is the Euclidean time
(call this the $x$-quantization).
Space is now the entire real line, $-\infty < \tau < \infty$.
The boundary condition is represented by a state
$|{\calB}\rangle$ inserted at $x=0$.
The correlation functions are expectation values of $x$-ordered
products of operators:
\eq
\expval{
\phi_{1}(\tau_{1},x_{1})\dots \phi_{n}(\tau_{n}, x_{n})}_{\calB}
=
\bra{0}
\phi_{1}(\tau_{1},x_{1})\dots \phi_{n}(\tau_{n}, x_{n})
\ket{\calB}
\en
where $\bra{0}$ is here the vacuum state,
and $x_{1} \ge x_{2} \ge \cdots \ge x_{n}$.
The correlation functions are normalized, $\langle 0 \ket{\calB}
=1$.

Suppose $Q$ is the generator of a symmetry of the bulk
system, specifically a global conformal symmetry.
Then $\bra{0} Q = 0$.
It seems reasonable to suppose that
\eq
\bra{0} Q
\phi_{1}(\tau_{1},x_{1})\dots \phi_{n}(\tau_{n}, x_{n})
\ket{\calB}
=0\,.
\en
For the bulk global conformal symmetry
group, $SL(2,{\mathbb C})$,
we can take $Q$ to
be any of the six generators
\eqa
Q_{n} &=& \int_{-\infty}^{\infty} d\tau \,
(x+iv\tau)^{n} T_R(x + iv\tau)\, ,
\nonumber \\
\bar Q_{n} &=& \int_{-\infty}^{\infty} d\tau \,
(x-iv\tau)^{n}\bar T_L(x-iv\tau) \qquad n=0,1,2 .
\ena
We should note that there is a subtlety in the above
reasoning.  A conserved charge $Q$ is defined as an integral
\eq
Q = \int\limits_{-\infty}^{+\infty}d\tau \, j^{x}(\tau,x)
= \lim_{R\to \infty} \int\limits_{|\tau|<R} d\tau
j^{x}(\tau,x)
\en
where $j^{x}(\tau,x)$ is the $x$-component of the
corresponding current.  Since we have very little knowledge
of the properties of the state $\ket{{\calB}}$ in general,
we can worry that the limit $R\to \infty$ taken in a
correlator
\eq
\lim_{R\to \infty} \bra{0}
\int\limits_{|\tau|<R} d\tau j^{x}(\tau,x)
\phi_{1}(\tau_{1},x_{1})\dots \phi_{n}(\tau_{n}, x_{n})
\ket{\calB}
\en
might not converge to zero.  The problem with this limit
could be due to a high density of low energy states present
in $\ket{{\calB}}$.  If for some reason the above limit
does not converge to zero this would mean that the
asymptotic symmetry $Q$ is spontaneously broken by the
boundary condition $|{\calB}\rangle$.  We will assume that
this does not happen or, in other words, the charge $Q$
exists and is an asymptotic symmetry in the theory on a half
plane with the given boundary condition $\ket{{\calB}}$.
The condition $\bra{0} Q = 0$, understood in the above sense,
implies that correlation functions are asymptotically
conformally invariant.
That is, correlation functions containing a commutator
$[Q,\phi(\tau,x)]$ asymptotically vanish
for $x\to \infty$.
But the condition $\bra{0} Q = 0$ is stronger.

At temperature $T>0$, the Euclidean time is compact, so
there is no subtlety in expressing the bulk conformal
invariance.  In Appendix~\ref{app:A} we use the bulk
conformal invariance at $T>0$, then take the $T\to 0$ limit,
and, assuming that dispersion relations behave in a
continuous fashion in this limit, we reproduce all the
consequences of the $\bra{0}Q=0$ assumption that we are
making here.  It cannot be considered as a derivation of the
$\bra{0}Q=0$ condition, though, because the assumption of
continuity at $T=0$ is essentially as strong as the
$\bra{0}Q=0$ condition itself.

With this assumption, we can write, for any $x>0$,
\eqa
0&=&\int\limits_{-\infty}^{+\infty}d\tau\,
(x+iv\tau)^{n}\langle T_R(x+iv\tau)\theta(0)\rangle_{c}
\qquad n=0,1,2 \nonumber\\
0 &=&
\int\limits_{-\infty}^{+\infty}d\tau\,
(x-iv\tau)^{n}\langle T_L(x-iv\tau)\theta(0)\rangle_{c}
\qquad n=0,1,2
\ena
or, equivalently,
\eqa
0&=&\int\limits_{-\infty}^{+\infty}d\tau\,
\tau^{n}\langle T_R(x+iv\tau)\theta(0)\rangle_{c}
 \qquad n=0,1,2 \nonumber\\
0&=&
\int\limits_{-\infty}^{+\infty}d\tau\,
\tau^{n}\langle T_L(x-iv\tau)\theta(0)\rangle_{c}
\qquad n=0,1,2
.
\label{eq:twopoint}
\ena

Now we consider the spectral representations for
the two-point correlation functions.
In Euclidean time we have
\begin{eqnarray}\label{tTEucl}
\langle T_R(x+iv\tau)\theta(0)\rangle_{c} &=&
\frac{1}{2\pi i}
\int\limits_{-\infty}^{\infty}d\omega\,
[\theta(\tau)\theta(\omega) - \theta(-\tau)\theta(-\omega)]
e^{-\omega(\tau-ix/v)}  A_{\theta R}(\omega) \nonumber\\
\langle T_L(x-iv\tau)\theta(0)\rangle_{c} &=&
\frac{1 }{2\pi i}
\int\limits_{-\infty}^{\infty}d\omega\,
[\theta(-\tau)\theta(\omega) - \theta(\tau)\theta(-\omega)]
e^{\omega(\tau+ix/v)}  A_{\theta L}(\omega) \\
\langle \theta(\tau)\theta(0)\rangle_{c} &=& \frac{1}{2\pi
}\int\limits_{0}^{\infty}d\omega\, e^{-\omega|\tau|}
A_{\theta\theta}(\omega)\, .
\end{eqnarray}
The boundary conservation equation (\ref{boundaryc}), written as
\eq
\partial_{\tau}\theta(\tau) =
-i [T_R(iv\tau) -   T_L(-iv\tau)] \,,
\en
implies\footnote{In deriving this equation
from the boundary conservation equation one uses the fact
that the
correlator $\langle \theta(\tau)\theta(\tau')\rangle_{c}$
vanishes for large separation and hence there cannot be a
term proportional to $\delta(\omega)$ in the spectral function}
\begin{equation}\label{spec}
A_{\theta\theta}(\omega) = \omega^{-1}(A_{\theta R}(\omega)
+ A_{\theta L}(-\omega))
=\omega^{-1}(A_{\theta R}(-\omega)
+ A_{\theta L}(\omega))
 \, .
\end{equation}
We note, though we do not use here, that
$T_R(x-vt)$ and $T_L(x+vt)$ are self-adjoint operators, so
\begin{equation} \label{parity}
\overline{A_{\theta R}(\omega)} = A_{\theta R}(-\omega)
\, , \quad \overline{A_{\theta L}(\omega)} = A_{\theta L}(-\omega) \, . 
\end{equation}
and, by reflection positivity,
$A_{\theta\theta}(\omega)\ge 0$.

The spectral functions $A_{\theta R}(\omega)$, $A_{\theta
L}(\omega)$ are related to the commutators as
\begin{eqnarray}
&&A_{\theta R}(\omega) = \int\limits_{-\infty}^{+\infty}dt
e^{i\omega t}\langle i[T_{R}(0,t), \theta(0)]\rangle =
\int\limits_{0}^{+\infty}dt e^{i\omega t}\langle
i[T_{R}(0,t), \theta(0)]\rangle \, , \nonumber \\
&& A_{\theta L}(\omega) = \int\limits_{-\infty}^{+\infty}dt
e^{-i\omega t} \langle -i[T_{L}(0,t),\theta(0)]\rangle =
\int_{-\infty}^{0}dt e^{-i\omega t} \langle
-i[T_{L}(0,t),\theta(0)]\rangle
\label{eq:Aanalyticity}
\end{eqnarray}
where the final forms of the equations
are consequences of the chirality of the energy-momentum
currents, equation (\ref{chirality}), and
causality (the vanishing of equal-time commutators at
nonzero separation).
It follows from the final forms of equations (\ref{eq:Aanalyticity})
that the spectral functions
$A_{\theta R}(\omega)$ and $A_{\theta L}(\omega)$
are analytic in the
upper half complex plane.
Again, we note but do not use here that energy conservation 
at the boundary combined with the bulk equal-time 
commutation relations of the chiral energy-momentum currents 
now imply $A_{\theta R}(\omega) = A_{\theta L}(\omega)$,
so $A_{\theta\theta}(\omega) = (2/\omega ) \Real\,A_{\theta 
R}(\omega)$.

The conformal invariance of the bulk vacuum
at large $x$, expressed by equations (\ref{eq:twopoint}),
is equivalent to
\begin{equation} \label{eq}
0=\int\limits_{-\infty}^{+\infty}d\omega\, e^{i\omega
x/v}\omega^{-n-1}A_{\theta R}(\omega) =
\int\limits_{-\infty}^{+\infty}d\omega\, e^{-i\omega
x/v}\omega^{-n-1}A_{\theta L}(\omega)\,, \quad n=0,1,2\, .
\end{equation}

It follows from (\ref{parity}) and (\ref{eq}) that
\begin{equation}\label{eq2}
0=\int\limits_{-\infty}^{+\infty}d\omega\,{\omega^{-3}}
[\sin(\omega x/v){ \rm \bf Re}A_{\theta R}(\omega) +
\cos(\omega x/v){ \rm \bf Im}A_{\theta R}(\omega)] \, .
\end{equation}
This implies that the functions ${ \rm \bf Re}A_{\theta R}(\omega)/\omega^{2}$ and
${ \rm \bf Im}A_{\theta R}(\omega)/\omega^{3}$ are integrable at $\omega=0$.
This implies in particular that
\begin{equation} \label{im0}
\lim_{\omega \to 0}\frac{{\rm \bf Im} A_{\theta R}(\omega)}{\omega^{2}}=0\, .
\end{equation}
Also, taking $x\to 0$ in (\ref{eq2}), we obtain a sum rule
\begin{equation}\label{sumrule0}
\int\limits_{-\infty}^{+\infty}d\omega\,
\frac{{ \rm \bf Im}A_{\theta R}(\omega)}{\omega^{3}} = 0\, .
\end{equation}

The just derived integrability properties of ${\rm \bf Im
}A_{\theta R}(\omega)$ and ${\rm \bf Re}A_{\theta
R}(\omega)$ at $\omega=0$, and canonical UV behavior, and
analyticity in the upper half plane allow one to write the
following subtracted dispersion relations
\begin{eqnarray} \label{disprel0}
\frac{{ \rm \bf Re}A_{\theta R}(\omega)}{\omega^{2}} &=&
\frac{1}{2\pi}\int\limits_{-\infty}^{+\infty}d\eta \,
\frac{{ \rm \bf Im}A_{\theta R}(\eta)}{\eta^{2}}{\cal
P}\left(\frac{1}{\eta - \omega}\right) \, , \nonumber \\
\frac{{ \rm \bf Im}A_{\theta R}(\omega)}{\omega^{2}} &=&
-\frac{1}{2\pi}\int\limits_{-\infty}^{+\infty}d\eta \,
\frac{{ \rm \bf Re}A_{\theta R}(\eta)}{\eta^{2}}{\cal
P}\left(\frac{1}{\eta - \omega}\right)
\end{eqnarray}
for $\omega\ne 0$.  The integrability of ${ \rm \bf
Im}A_{\theta R}(\omega)/\omega^{3}$ at $\omega=0$ allows us
to take the limit $\omega \to 0$ of the first dispersion
relation in (\ref{disprel0}) in a straightforward way and we
obtain
$$
\lim\limits_{\omega \to 0}\frac{{ \rm \bf Re}A_{\theta
R}(\omega)}{\omega^{2}} =
\frac{1}{2\pi}\int\limits_{-\infty}^{+\infty}d\eta\, \frac{{
\rm \bf Im}A_{\theta R}(\eta)}{\eta^{3}} \, .
$$
This equation together with the  sum rule (\ref{sumrule0}) and equation (\ref{im0})
 imply that   
\begin{equation} \label{tT''}
\lim_{\omega \to 0}\frac{A_{\theta R}(\omega)}{\omega^{2}}=0\, .
\end{equation}
By the same argument,
\begin{equation} \label{tT'''}
\lim_{\omega \to 0}\frac{A_{\theta L}(\omega)}{\omega^{2}}=0\, .
\end{equation}
Therefore by (\ref{spec})
\begin{equation}
\lim_{\omega\to 0} \frac{A_{\theta \theta}(\omega)}{\omega} = 0
\end{equation}
which in turn implies (\ref{lim}), which was to be shown.

Noting that $\theta(\tau)$ has a canonical scaling dimension
1, we infer that in the infrared limit $\mu \to \infty$ the
two-point function at hand goes to zero:
\eq
\lim_{\mu \to
\infty}\langle \theta(\tau)\theta(0)\rangle_{c} = 0 \, .
\en
In a quantum field theory, a local field with vanishing
two-point function annihilates
the ground state, and therefore has vanishing correlation
functions with all other fields.
Thus, if we can assume that we obtain a local
boundary quantum field theory in the infrared
limit $\mu\to\infty$,
and if we can assume that
$$
\bra{0} Q
\phi_{1}(\tau_{1},x_{1})\dots \phi_{n}(\tau_{n}, x_{n})
\ket{\calB} = 0
$$
for all the bulk global conformal symmetries $Q$ acting far
from the boundary,
then we can conclude that the limiting theory in the infrared
has to be conformal, with a finite boundary entropy.  In such
cases (when locality is preserved all the way to the far
infrared) the boundary entropy stays bounded from below.

\section{Proof of the gradient formula in the real time formalism}\label{real_time}
Here, we will use the machinery of real time spectral
analysis for equilibrium boundary quantum field theory in
$1+1$ dimensions, as developed in \cite{FriedanFlowB}.
Using the real time formalism, we will re-state the proof
that $\dif s/\dif T \ge 0$ and the proof of the gradient
formula for the boundary beta-function, $\partial s/\partial
\lambda^{a}=-g_{ab}\beta^{b}$.  The Riemannian metric on the
space of boundary couplings, $g_{ab}(\lambda)$, is
$\chi_{ab}(0)/T$, the renormalized static susceptibility
matrix of the boundary, divided by temperature.  The dynamic
susceptibility matrix of the boundary, $\chi_{ab}(\omega)$,
is renormalized by natural subtractions in such a way that the
static susceptibility matrix, $\chi_{ab}(0)$, remains
positive.

The first step will be to show that
\eq
\frac{\dif s}{\dif T} = \frac12\,
T^{-2} \Imag\,  F'(0)
\label{eq:basic_formulaA}
\en
where
\eq
F(\omega) =
\int_{-\infty}^{\infty} \dif t\,\,
\me^{-i\omega t} \expvalequil{i
\comm{\TL(vt)+\TR(vt)}{\theta(0)}}
\:.
\label{eq:basic_formulaB}
\en
This is a Kubo formula for the change in the boundary
entropy in response to a local change in the temperature at
the boundary.  The response function $F(\omega)$ is analytic
in the upper half-plane.  On the real axis, $-\Real\,
F(\omega) \ge 0$.

The second step is to show that $\Imag\, F'(0) \ge 0$, and
therefore $\dif s/\dif T \ge 0$, by deriving a dispersion
formula for $\Imag\,F(\omega)$ in terms of
$\Real\,F(\omega)$.  The naive, unsubtracted dispersion
formula is divergent, because $F(\omega)$ can grow as fast
as $\omega$ for large $\omega$, by canonical dimensional
analysis in the ultraviolet limit.  Fortunately, bulk
conformal invariance will imply a vanishing formula,
$F(i2\pi T)=0$, which gives a natural subtraction
point.  The subtracted dispersion formula converges as long
as the ultraviolet behavior is canonical (as long as the
system approaches a renormalization group fixed point in the
ultraviolet).  The subtracted dispersion formula will still
imply $\Imag\, F'(0) \ge 0$, and therefore $\dif s/\dif T >0$.

Equations~(\ref{eq:basic_formulaA})-(\ref{eq:basic_formulaB})
were derived in Ref.~\cite{FriedanFlowB} by considering the
flow of entropy in real time, in and out of the boundary, in
analogy with the flow of electric charge in an electric
circuit.  The flow of entropy is described by an entropy
current operator, which is just the energy current divided
by the temperature.  The right-moving entropy current
operator is the right-moving energy current divided by
temperature, $j_{L}(x,t)=\TL(x,t)/T$.  The boundary entropy
``charge'' operator is $q_{S}(t)=-\theta(t)/T$.  The Kubo
formula for the entropic ``admittance'' of the boundary was
written, using the chirality of the bulk entropy currents,
\eq
Y_{S}(\omega)
=
\int_{-\infty}^{\infty} \dif t\,\,
\me^{-i\omega t} \expvalequil{i
\comm{ j_{L}(0,t)}{- q_{S}(0)}}
\:.
\en
The entropic ``capacitance'' of the boundary is
\eq
\frac{\dif s}{\dif T} = \lim_{\omega\rightarrow 0}
\frac1{i\omega} Y_{S}(\omega)= \Imag Y'_{S}(0) .
\en
These are exactly
equations~(\ref{eq:basic_formulaA})-(\ref{eq:basic_formulaB}),
since $T^{-2}F(\omega)/2 =Y_{S}(\omega)$.  Here, we derive
equations~(\ref{eq:basic_formulaA})-(\ref{eq:basic_formulaB})
directly.

The proof is based on the following assumptions:
\begin{enumerate}
\item There is a local, symmetric energy-momentum tensor
$T^{\mu}_{\nu}(x,t)$, $T_{\mu\nu}=T_{\nu\mu}$.
\item The system is locally scale invariant in the bulk.
The trace of the energy-momentum tensor vanishes in the
bulk, $T_{\mu}^{\mu}=0$.
\item The bulk system is conformally invariant.  The bulk
ground state in the ``closed'' channel is annihilated by the
Virasoro operators $L_{0}+\bar L_{0}$ and
$L_{1}+\bar L_{1}$.
\item The system exhibits canonical scaling behavior in the
ultraviolet (goes to a renormalization group fixed
point in the ultraviolet).
\item The system is in equilibrium at temperature $T$.
Equilibrium expectation values of commutators of local operators,
$\expvalequil{\comm{\mathcal{O}_{1}(t_{1})}{\mathcal{O}_{2}(t_{2})}}$,
go to zero at large times, $t_{1}-t_{2}\rightarrow \pm\infty$.
\item The Fourier transforms
$$
\int \dif t\,\,\me^{-i\omega t}
\expvalequil{\comm{\mathcal{O}_{1}(t_{1})}{\mathcal{O}_{2}(t_{2})}}
$$
are smooth functions of the frequency $\omega$,
for any local operators $\mathcal{O}_{1}(t)$, $\mathcal{O}_{2}(t)$.
\end{enumerate}

\subsection{The Kubo formula for $\dif s/\dif T$}
The bulk energy density operator is
$
[v^{-1}\TR(x-vt)+ v^{-1}\TL(x+vt)]
$
and the boundary energy operator is $-\theta(t)$,
so
the thermodynamic energy of the full system (of length $L$) is
$$
-\partialby{\beta} \ln Z = \expvalequil{\Hzero}
=  \expvalequil{-\theta(t) + \int_{0}^{L} \dif x \,\,
[v^{-1}\TR(x-vt)+ v^{-1}\TL(x+vt)]}
\:.
$$
The equilibrium expectation values
$\expvalequil{\TR(x-vt)}$ and $\expvalequil{\TL(x+vt)}$
are constant in $x$ because they are
independent of time,
so $\expvalequil{v^{-1}\TR(x-vt)+
v^{-1}\TL(x+vt)}$ is the bulk energy density,
$c\pi L/6\beta^{2}$, which is
determined by bulk conformal invariance
up to the value of the bulk conformal central charge, $c$.
The difference between the total thermodynamic energy
and the bulk energy
is the thermodynamic boundary energy:
$$
-\partialby{\beta} \ln \zboundary = \expvalequil{-\theta(t)}
\:.
$$

The boundary entropy is given by formula (\ref{s}). Thus we have
\eq
T^{2} \frac{\partial s}{\partial T}
= -\frac{\partial s}{\partial \beta}
= \beta \partialby\beta \expvalequil{\theta(0)}
=-\beta \expvalc{\Hzero\, \theta(0)}
\:.
\en
Approximate the Hamiltonian by
introducing an arbitrary cutoff point $x_{1}>0$:
$$
H(x_{1},t) = -\theta(t) +
\int_{0}^{x_{1}}\dif x \,\,
[v^{-1}\TR(x-vt)+ v^{-1}\TL(x+vt)]\, .
$$
Approximate $\beta \Hzero$ by integrating over imaginary
time, $\tau = i t$, from $0$ to $\beta$:
$$
\beta \Hzero
\approx \int_{0}^{- i  \beta}\dif t \,\, i H(x_{1},t)
\:.
$$
Then
$$
T^{2} \frac{\partial s}{\partial T} =
\lim_{x_{1}\rightarrow \infty}
\int_{0}^{- i \beta}\dif t \,\, (-i)
\expvalc{H(x_{1},t)\, \theta(0)}
\:.
$$
In fact, there is no dependence on $x_{1}$, because
\begin{eqnarray*}
\lefteqn{
\partialby{x_{1}}
\int_{0}^{- i  \beta}\dif t \,\,
\expvalc{H(x_{1},t)\, \theta(0)}
}\hspace{2em}  \\
&&
=
\int_{0}^{- i  \beta}\dif t \,\,
\expvalc{[v^{-1}\TR(x_{1}-vt)+ v^{-1}\TL(x_{1}+vt)]\,
\theta(0)}
\end{eqnarray*}
which is zero because the rhs, evaluated in the ``closed''
channel where $x$ is imaginary time, is a matrix element of
the Virasoro operator $L_{0}+\bar L_{0}$ between a boundary
state and the bulk ground state, and the bulk ground state
is annihilated by $L_{0}+\bar L_{0}$.  Therefore, for any
$x_{1}>0$,
\eq
T^{2}\frac{\partial s}{\partial T} =
\int_{0}^{- i \beta}\dif t \,\, (-i)
\expvalc{H(x_{1},t)\, \theta(0)}
\:.
\label{eq:eucltime}
\en
Now deform the contour of integration in the
standard way to obtain the Kubo formula:
\eqa
\frac{\partial s}{\partial T}
&=&
T^{-2}
\left ( \int_{-\infty- i  \beta}^{0- i\beta}
-
\int_{-\infty}^{0}
\right )
\dif t \,\, (-i)
\expvalc{H(x_{1},t)\, \theta(0)}\nonumber \\
&=&
T^{-2}
\int_{-\infty}^{0}
\dif t \,\,
\expvalequil{i\comm{H(x_{1},t)}{\theta(0)}}
\:.
\ena
This is the Kubo formula for the entropic ``capacitance'' of
the boundary, which was derived in \cite{FriedanFlowB} as
the infinitesimal change in the entropic
``charge,'' $-\theta(t)/T$, produced in real time by an
infinitesimal change in the entropic ``potential'' of the
boundary.

The integrand in the Kubo formula is a distribution in $t$.
In the derivation, the contour deformation in the complex
time plane is justified by Gauss' law for distributions,
applied on the region $-i\beta \le\Imag\, t \le 0$,
$\Real\, t \le 0$.  The integral over the boundary of this
region vanishes.  The boundary integral
can be separated unambiguously into two parts --- the
integral over the imaginary $t$ axis from $0$ to
$-i\beta$, and the rest --- because the
integrand is an ordinary function near $t=0$ and near
$t=-i \beta$.  In general, equilibrium expectation values
satisfy
$$
\expvalequil{H(x_{1},t-i \beta)\,\theta(0)}
=\expvalequil{\theta(0)\,H(x_{1},t)}
$$
but here,
$$
\expvalequil{\comm{H(x_{1},t)}{\theta(0)}}
= \expvalequil{\comm{\Hzero}{\theta(0)}}
=0
$$
for all real $t$ in the range $-x_{1}<vt<x_{1}$, by
causality.  It takes at least time $x_{1}/v$ for any effect
of the cutoff at $x_{1}$ to reach the boundary, or vice
versa.  Therefore the integrand in the Kubo formula is
identically zero near $t=0$.  The equilibrium correlation
function, $\expvalequil{H(x_{1},t)\,\theta(0)}$, is periodic
on the imaginary $t$ axis, with period $i \beta$,
without singularity at $t=0$ or $t= i \beta$.

A second Kubo formula is obtained by deforming the
integration contour in equation (\ref{eq:eucltime})
to positive times:
\eq
\frac{\partial s}{\partial T}=
T^{-2}\int_{0}^{\infty}
\dif t \,\,
\expvalequil{-i\comm{H(x_{1},t)}{\theta(0)}}
\:.
\label{eq:secondKubo}
\en

\subsection{Using chirality of the energy currents}
Local conservation of energy implies
$$
\partial_{t} H(x_{1},t)
=
-\TR(x_{1}-vt)+ \TL(x_{1}+vt)
\:,
$$
thus
\eq
T^{2}
\frac{\partial s}{\partial T} =
\int_{-\infty}^{0}
\dif t \,\,
\int_{-\infty}^{t} \dif t' \,\,
\expvalequil{i\comm{-\TR(x_{1}-vt')+\TL(x_{1}+vt')}
{\theta(0)}}
\:.
\en
The boundary term at $t'=-\infty$ can be neglected because
equilibrium expectation values of commutators of local
operators decay to zero at large times.  Now use the
chirality of the bulk energy currents.  For all
$t'<x_{1}/v$,
$$
\expvalequil{\comm{\TR(x_{1}-vt')}{\theta(0)}}
=
\expvalequil{\comm{\TR(x_{1}-vt',0)}{\theta(0)}}
=0
$$
as an equal-time commutator of spatially separated
operators.
Therefore
\eqa
T^{2} \frac{\partial s}{\partial T}
&=&
\int_{-\infty}^{0}
\dif t \,\,
\int_{-\infty}^{t} \dif t' \,\,
\expvalequil{i \comm{\TL(x_{1}+vt')}{\theta(0)}}\nonumber\\
&=&
\int_{-\infty}^{0}
\dif t' \,\,
\int_{t'}^{0} \dif t \,\,
\expvalequil{i \comm{\TL(x_{1}+vt')}{\theta(0)}}\nonumber\\
&=&
\int_{-\infty}^{0}
\dif t' \,\,
(-t')
\expvalequil{i \comm{\TL(x_{1}+vt')}{\theta(0)}}\nonumber \\
&=&
\int_{-\infty}^{\infty}
\dif t' \,\,
(-t')
\expvalequil{i \comm{\TL(x_{1}+vt')}{\theta(0)}}
\:.
\ena
In the last step,
it makes no difference to
extend the time integral to $+\infty$,
because, for all $t'>- x_{1}/v$,
$$
\expvalequil{\comm{\TL(x_{1}+vt')}
{\theta(0)}}
=
\expvalequil{\comm{\TL(x_{1}+vt',0)}
{\theta(0)}}
=0
$$
as an equal-time commutator of spatially
separated operators.
Next,
change the integration variable to $t=t'-x_{1}/v$, obtaining
$$
T^{2} \frac{\partial s}{\partial T} =
\int_{-\infty}^{\infty}
\dif t \,\,
(-t+x_{1}/v)
\expvalequil{i \comm{\TL(vt)}{\theta(0)}}
\:.
$$
The term proportional to $x_{1}$ vanishes by (\ref{H2})
thus
\eq \label{eq:HzeroTL}
T^{2} \frac{\partial s}{\partial T} =
\int_{-\infty}^{\infty}
\dif t \,\,
(-t)\,
\expvalequil{i \comm{\TL(0,t)}{\theta(0)}}
\:.
\en
In terms of the response function
\eq
F_{L}(\omega)  = \int_{-\infty}^{\infty} \dif t\,\,
\me^{-i\omega t} \expvalequil{i
\comm{ \TL(0,t)}{ \theta(0)}}
\:,
\label{eq:FL}
\en
\eq
T^{2} \frac{\partial s}{\partial T} = i^{-1} F'_{L}(0)
\:.
\en
$F_{L}(\omega)$ is analytic in the upper half-plane because
the commutator vanishes for all $t>0$,
by the chirality of the energy current.

By similar reasoning, the second Kubo formula,
equation (\ref{eq:secondKubo}), becomes
\eq
T^{2} \frac{\partial s}{\partial T} =
\int_{-\infty}^{\infty}\dif t
\,\,
t \,
\expvalequil{-i\comm{\TR(0,t)}{\theta(0)}}
\:.
\en
In terms of the response function
\eq
F_{R}(\omega)  = \int_{-\infty}^{\infty} \dif t\,\,
\me^{i\omega t} \expvalequil{-i
\comm{ \TR(0,t)}{ \theta(0)}}
\:,
\label{eq:FR}
\en
\eq
T^{2} \frac{\partial s}{\partial T} = i^{-1} F'_{R}(0)
\:.
\en
$F_{R}(\omega)$ is analytic in the upper half-plane because
the commutator vanishes for all $t<0$, by the chirality of
the energy current.  Finally, define
\eqa
F(\omega) &=& F_{L}(\omega) +  F_{R}(\omega) \nonumber \\
&=& \int_{-\infty}^{\infty} \dif t\,\,
\me^{-i\omega t} \expvalequil{i
\comm{ \TL(vt)-\TR(vt)}{\theta(0)}}
\ena
so
\eq \label{dsdt}
T^{2} \frac{\partial s}{\partial T} = \frac12 i^{-1} F'(0)
\:.
\en

\subsection{Properties of $F(\omega)$}\label{propF}
\begin{enumerate}
\item $F(\omega)$ is analytic in the upper half-plane.
\item $\overline{F(\omega)} = F(-\bar\omega)$.
\label{prop:reality}
\item $F(0)=0$.
\item $T^{2} \frac{\partial s}{\partial T} = \frac12 \Imag\,
F'(0)$.
\item $F(\omega)/\omega^{2} \rightarrow 0$ as
$\omega\rightarrow \pm\infty$.
\item \label{eq:prop_realF}
$\Real\,F(\omega) = \int_{-\infty}^{\infty} \dif t\,\,
\me^{-i\omega t} \omega
\expvalequil{\comm{\theta(t)}{ \theta(0)}}
$.
\item \label{prop:positivity}
$-\Real\,F(\omega)\ge 0$ for all real $\omega$.
\item \label{prop:vanishing}
$F(i 2\pi  T) = 0$.
\end{enumerate}

$F(\omega)$ is analytic in the upper half-plane because both
$F_{L}(\omega)$ and $F_{R}(\omega)$ are.
$\overline{F(\omega)} = F(-\bar \omega)$ because
$\TL(x,t)$, $\TR(x,t)$ and $\theta(t)$ are self-adjoint
operators.
Property~\ref{prop:reality} implies that $F'(0)$ is
imaginary, so
$T^{2} {\partial s}/{\partial T} = \Imag\, F'(0)/2$.
$F(0)=0$ by equation (\ref{eq:HzeroTL}).
$F(\omega)/\omega^{2} \rightarrow 0$ as
$\omega\rightarrow \infty$ by
canonical dimensional analysis in the ultraviolet limit.
$\TL(x,t)$ and $\TR(x,t)$ each has scaling dimension $2$, while
$\theta(t)\dif t=\beta^{a}\phi_{a}(t)\dif t$ goes
to zero in the ultraviolet limit,
assuming that the system goes to a renormalization group
fixed point in the ultraviolet.
For property~\ref{eq:prop_realF},
evaluate
\begin{eqnarray*}
\Real\,F(\omega) &=& \Real\,
\int_{-\infty}^{\infty} \dif t\,\,
\me^{-i\omega t} \expvalequil{i
\comm{\TL(0,t)-\TR(0,t)}{ \theta(0)}} \\
&=& \Real\,
\int_{-\infty}^{\infty} \dif t\,\,
\me^{-i\omega t} \expvalequil{i
\comm{-\theta'(t)}{ \theta(0)}} \\
&=& \int_{-\infty}^{\infty} \dif t\,\,
\me^{-i\omega t} \omega
\expvalequil{\comm{\theta(t)}{ \theta(0)}}
\:.
\end{eqnarray*}
For property~\ref{prop:positivity},
define the operator Fourier modes
$$
\tilde\theta(\omega) = \int_{-\infty}^{\infty} \dif t \,\,
\me^{i\omega t} \theta(t)
$$
satisfying
$$
\tilde\theta(\omega)^{\dagger}= \tilde\theta(-\omega)
$$
$$
\comm{\Hzero}{\tilde\theta(\omega)} =
- \omega \tilde\theta(\omega)
\:.
$$
Then we have
\eq
2\pi \delta(\omega'+\omega)
\Real\, \omega^{-1}F(\omega)
=
\expvalequil{\comm{\tilde\theta(\omega')}{\tilde\theta(\omega)}}
\:.
\en
Next note that
$$
\expvalequil{
\comm{\tilde\theta(\omega')}{\tilde\theta(\omega)}}
=
\left ( 1-\me^{\beta\omega} \right )
\expvalequil{\tilde\theta(\omega')\,\tilde\theta(\omega)}
$$
thus
$$
2\pi \delta(\omega'+\omega)
\Real\, \omega^{-1}F(\omega)
=
\left ( 1-\me^{\beta\omega} \right )
\expvalequil{\tilde\theta(\omega')\,\tilde\theta(\omega)}
\label{eq:thetatilde}
$$
which implies
\eq
-\Real F(\omega) \ge 0
\:.
\en
Finally, for property~\ref{prop:vanishing}, write
$$
\expvalequil{\comm{\TL(0,t)}{\theta(0)}} =
\frac1{2\pi}\int \dif \omega \,\,
\expvalequil{\comm{\TL(0,t)}{\tilde\theta(\omega)}}
$$
so
\begin{eqnarray*}
\expvalequil{\comm{\TL(0,t)}{\tilde\theta(\omega)}}
&=& \me^{i\omega t} (-i) F_{L}(\omega)\\
\left ( 1-\me^{\beta\omega} \right )
\expval{\TL(0,t)\, \tilde\theta(\omega)}
&=&
\expvalequil{\comm{\TL(0,t)}{\tilde\theta(\omega)}} \\
\expvalc{\TL(0,t)\, \tilde\theta(\omega)}
&=&
\me^{i\omega t}
\left ( 1-\me^{\beta\omega} \right )^{-1}
(-i) F_{L}(\omega)
\end{eqnarray*}
and therefore
\eq
\expvalc{\TL(0,t)\, \theta(0)}
=
\frac1{2\pi i} \int \dif \omega \,\,\me^{i\omega t}
\frac{ F_{L}(\omega)}{1-\me^{\beta\omega}}\, .
\en
We next Wick-rotate to imaginary time $\tau=it$, for
$0<\tau<\beta$, to get
\eq
\expvalc{\TL(0,-i\tau)\, \theta(0)}
=
\frac1{2\pi i} \int \dif \omega \,\,\me^{\omega \tau}
\frac{ F_{L}(\omega)}{1-\me^{\beta\omega}}
\en
Deform the contour of integration into the upper half-plane,
picking up the residues at the thermal poles:
\eq
\expvalc{\TL(0,-i\tau)\, \theta(0)}
=
\frac{-1}\beta \sum_{n=1}^{\infty}
\me^{i\omega_{n} \tau}
F_{L}(i\omega_{n})
\en
where
$$
\omega_{n} = \frac {2\pi n}{\beta}\, .
$$
Then, by chirality of the energy current,
\eq
\expvalc{\TL(x-iv\tau)\, \theta(0)}
=
\frac{-1}\beta \sum_{n=1}^{\infty}
\me^{-\omega_{n}  (x-iv\tau)/v}
F_{L}(i\omega_{n})
\:.
\en
Similarly, also for $0<\tau< \beta$,
\eq\label{FRcorr}
\expvalc{\TR(0,-i\tau)\, \theta(0)}
=
\frac1{2\pi i} \int \dif \omega \,\,\me^{-\omega \tau}
\frac{ F_{R}(\omega)}{\me^{-\beta\omega}-1}
\en
so
\eq
\expvalc{\TR(x+iv\tau)\, \theta(0)}
=
\frac{-1}\beta \sum_{n=1}^{\infty}
\me^{-\omega_{n}  (x+iv\tau)/v}
F_{R}(i\omega_{n})
\:.
\en
Setting $n=1$,
\eq
F(i\omega_{1}) =
\frac{2\pi}{ \beta} \expvalequil{K_{1}(x)\,\theta(0)}
\en
where
\eq
K_{1}(x) =
\frac{-\beta}{2\pi v}
\int_{0}^{ v\beta}\dif y \,\,
\left [
\me^{\omega_{1}  (x-iy)/v} T_{L}(x-iy)
+
\me^{\omega_{1}  (x+iy)/v} T_{R}(x+iy)
\right ]
\:.
\en
In the ``closed'' channel,
where $x$ is imaginary time,
$K_{1}(x)$ is
the Virasoro operator $L_{1}+\bar L_{1}$.
Therefore, in the ``closed'' channel,
$F(i\omega_{1})$ is a matrix element of
$L_{1}+\bar L_{1}$ between a boundary state and the bulk
ground state.
Global conformal invariance of the bulk system
implies that $L_{1}+\bar L_{1}$ annihilates the bulk ground
state in the ``closed'' channel.
Therefore $F(i\omega_{1})=F({2\pi i}/{\beta}) = 0$.

\subsection{Subtracted dispersion formula for $\Imag\,F'(0)$}
The vanishing formulas,
$F(0)=F({2\pi i}/{\beta})=0$,
allow writing
$$
\eta^{-1}F(\eta) = \frac1{2\pi i} \int_{-\infty}^{\infty} \dif
\omega\,\,
\left [
\frac1{\omega-\eta-i\epsilon}
- \frac{\omega+\eta+i\epsilon}{\omega^{2}+\omega_{1}^{2}}
\right ] \omega^{-1}F(\omega)
\:.
$$
The integral converges, because
$F(\omega)/\omega^{2}\rightarrow 0$ when $\omega\rightarrow
\pm \infty$.
Take the imaginary part to get the dispersion formula
\eq
\Imag \, \eta^{-1}F(\eta) =
\frac1{\pi} \int_{-\infty}^{\infty} \dif
\omega\,\,
\left [
\mathcal{P}\left ( \frac1{\omega-\eta}\right )
- \frac{\omega+\eta}{\omega^{2}+\omega_{1}^{2}}
\right ] \left [-\omega^{-1}\Real\,F(\omega)\right ]
\:.
\en
Take $\eta\rightarrow 0$ to get
\eq \label{disprel}
\Imag \, F'(0) =
\frac1{\pi} \int_{-\infty}^{\infty} \dif
\omega\,\,
\frac{-\Real\,F(\omega) }
{\omega^{2}(1+\omega_{1}^{-2}\omega^{2})}
\:.
\en
Thus $\Imag \, F'(0)\ge 0$.  Equality, $\Imag \, F'(0)= 0$, is
possible only if $\Real\,F(\omega)= 0$, which implies
$F(\omega)= 0$. It follows then from
equation (\ref{eq:thetatilde}) that $\tilde \theta(\omega)$
is proportional to $\delta(\omega)$, which implies that
$\theta(t)$ is a multiple of the identity. This means
that the boundary field theory is scale invariant.
Therefore $\partial s/\partial T \ge 0$,
with equality if and only if the boundary field theory is scale
invariant.

\subsection{The gradient formula}

Calculate
\eqa
\partial_{a} s
&=& \partial_{a} \left ( 1-\beta \partialby{\beta} \right ) z \nonumber \\
&=& \left ( 1-\beta \partialby{\beta} \right ) \partial_{a} z \nonumber \\
&=& \left ( 1-\beta \partialby{\beta} \right )
\expvalequil{\beta \phi_{a}(0)} \nonumber \\
&=& -\beta^{2} \partialby{\beta}\expvalequil{\phi_{a}(0)} \nonumber \\
&=& \beta^{2} \expvalc{\Hzero \, \phi_{a}(0)} \nonumber \\
&=& \beta
\int_{0}^{- i  \beta}\dif t \,\,i
\expvalc{H(x_{1},t) \, \phi_{a}(0)}
\ena
where the last expression is independent of $x_{1}$
and the result is thus exact,
as before.

Deform the integration contour to negative times to get
\eqa
-T \partial_{a} s
&=&
\left (
\int_{-\infty}^{0}
-
\int_{-\infty- i  \beta}^{0- i \beta}
\right )
\dif t \,\,i
\expvalc{H(x_{1},t) \, \phi_{a}(0)} \nonumber \\
&=&
\int_{-\infty}^{0}
\dif t \,\,
\expvalc{i \comm{H(x_{1},t)}{\phi_{a}(0)}} \nonumber \\
&=&
\int_{-\infty}^{0}
\dif t \,\,
\int_{-\infty}^{t}
\dif t' \,\,
\expvalc{i
\comm{\TL(x_{1},t')-\TR(x_{1},t')}{\phi_{a}(0)}} \nonumber \\
&=&
\int_{-\infty}^{0}
\dif t \,\,
(-t)
\expvalc{i
\comm{\TL(x_{1},t)}{\phi_{a}(0)}} \nonumber \\
&=&  i^{-1} F'_{La}(0)
\ena
where
\eq
F_{La}(\omega)  = \int_{-\infty}^{\infty} \dif t\,\,
\me^{-i\omega t} \expvalequil{i
\comm{ \TL(0,t)}{\phi_{a}(0)}}
\:.
\label{eq:Kubo}
\en
The spectral function $F_{La}(\omega)$ is analytic in the
upper half-plane.  Equation~(\ref{eq:Kubo}) is a Kubo
formula giving the response of the boundary fields to a
local change of temperature.

A second Kubo formula is obtained similarly by
deforming the contour to positive times:
\begin{equation}
-T \partial_{a} s = i^{-1} F'_{Ra}(0)
\end{equation}
where
\eq
F_{Ra}(\omega)  = \int_{-\infty}^{\infty} \dif t\,\,
\me^{i\omega t} \expvalequil{(-i)
\comm{ \TR(0,t)}{\phi_{a}(0)}}
\:.
\en
$F_{Ra}(\omega)$ is analytic in the upper half-plane.
Define
\eqa
F_{a}(\omega) &=& F_{La}(\omega) + F_{Ra}(\omega) \nonumber \\
&=& \int_{-\infty}^{\infty} \dif t\,\,
\me^{-i\omega t} \expvalequil{i
\comm{ \TL(vt)-\TR(vt)}{\phi_{a}(0)}}
\ena
so
\eq
-T \partial_{a} s = \frac12  i^{-1} F'_{a}(0)
\:.
\en
$F_{a}(\omega)$ satisfies
\begin{enumerate}
\item \label{prop:a_anal} $F_{a}(\omega)$ is analytic in the upper half-plane.
\item
$\overline{F_{a}(\omega)} = F_{a}(-\bar\omega)$.
\item
$F_{a}(0)=0$.
\item
$-T\partial_{a} s = \frac12 \Imag F'_{a}(0)$.
\item \label{prop:a_conf} $F_{a}(i 2\pi  T) = 0$.
\item \label{prop:a_beta} $\beta^{a}F_{a}(\omega)=F(\omega)$.
\item \label{prop:a_real}
$\Real\,F_{a}(\omega) = \Real\,\int_{-\infty}^{\infty} \dif t\,\,
\me^{-i\omega t} \omega
\expvalequil{\comm{\theta(t)}{ \phi_{a}(0)}}
$.
\item \label{prop:a_asympt}
$\Real\, F_{a}(\omega)/\omega^{2}\rightarrow 0$ as
$\omega\rightarrow \infty$.
\end{enumerate}
Properties~\ref{prop:a_anal}-\ref{prop:a_conf} are derived
just as for $F(\omega)$.
Property~\ref{prop:a_beta} follows from
$\theta(t) = \beta^{a} \phi_{a}(t)$.
For property~\ref{prop:a_real}, evaluate
\eqa
\Real\, F_{a}(\omega) &=&
\Real\,
\int_{-\infty}^{\infty} \dif t\,\,
\me^{-i\omega t} \expvalequil{i
\comm{
\TL(0,t)-\TR(0,t)}{\phi_{a}(0)}} \nonumber \\
&=&
\Real\,
\int_{-\infty}^{\infty} \dif t\,\,
\me^{-i\omega t} \expvalequil{i
\comm{ -\theta'(t)}{\phi_{a}(0)}} \nonumber \\
&=&
\Real\,
\int_{-\infty}^{\infty} \dif t\,\,
\me^{-i\omega t}
\omega
\expvalequil{\comm{ \theta(t)}{\phi_{a}(0)}}
\ena
Property~\ref{prop:a_asympt} follows from
property~\ref{prop:a_real} and canonical scaling in the
ultraviolet, since the boundary operators, $\phi_{a}(t)$,
all have scaling dimensions $\le 1$,
and $\theta(t)\dif t$ vanishes in the ultraviolet limit.

As before, the vanishing formulae allow writing a
subtracted dispersion formula:
\eq
\Imag \, \eta^{-1}F_{a}(\eta) =
\frac1{\pi} \int_{-\infty}^{\infty} \dif
\omega\,\,
\left [
\mathcal{P}\left ( \frac1{\omega-\eta}\right )
- \frac{\omega+\eta}{\omega^{2}+\omega_{1}^{2}}
\right ] \left [-\omega^{-1}\Real\,F_{a}(\omega)\right ]
\en
where $\omega_{1}=2\pi T$.
Take $\eta\rightarrow 0$ to get
\eq
\Imag \, F'_{a}(0) =
\frac1{\pi} \int_{-\infty}^{\infty} \dif
\omega\,\,
\frac{-\Real\,F_{a}(\omega)}
{\omega^{2}(1+\omega_{1}^{-2}\omega^{2})}
\en
Now use property~\ref{prop:a_real} and the identity
$\theta(t)=\beta^{a}\phi_{a}(t)$ to write
\eq
- \Real\, F_{a}(\omega)
=
\Real\,\omega G_{ab}(\omega) \beta^{b}
\en
where
\eq
G_{ab}(\omega) =
\int_{-\infty}^{\infty} \dif t\,\,
\me^{-i\omega t}
\expvalequil{\comm{ -\phi_{b}(t)}{\phi_{a}(0)}}
\:.
\en
In terms of the operator Fourier modes
$$
\tilde\phi_{a}(\omega) = \int_{-\infty}^{\infty} \dif t \,\,
\me^{i\omega t} \phi_{a}(t)
\:,
$$
$$
\expvalequil{\comm{-\tilde\phi_{b}(\omega')}{\tilde\phi_{a}(\omega)}}
= 2\pi  \delta(\omega'+\omega)
G_{ab}(\omega)
$$
so
\eq
\expvalc{\tilde\phi_{b}(\omega')\,\tilde\phi_{a}(\omega)}
= 2\pi  \delta(\omega'+\omega)
\frac{G_{ab}(\omega)}
{\me^{\beta\omega}-1} \, .
\label{eq:Gabtildephi}
\en
Therefore
\begin{enumerate}
\item $G_{ab}(\omega)/\omega$ is a nonnegative hermitian matrix,
\item $ G_{ab}(-\omega) = - \overline{G_{ab}(\omega)}$,
\item $G_{ab}(-\omega)=-G_{ba}(\omega)$.
\end{enumerate}
The dispersion formula for $\Imag F'_{a}(0)$ becomes
\eq
\Imag F'_{a}(0)
=
\frac1{\pi} \int_{-\infty}^{\infty} \dif
\omega\,\,
\frac{\Real\,G_{ab}(\omega) \beta^{b}}
{\omega (1+\omega_{1}^{-2}\omega^{2})}
\en
which is the gradient formula,
$$
\partial_{a} s = - g_{ab} \beta^{b}
$$
with
\eq
T g_{ab} =
\frac{1}{2\pi} \int_{-\infty}^{\infty} \dif
\omega\,\,
\frac{\Real\,G_{ab}(\omega)}
{\omega(1+\omega_{1}^{-2}\omega^{2})}
\label{eq:gab}
\en
being a positive symmetric matrix, a Riemannian metric on the
space of boundary couplings.

\subsection{The renormalized boundary susceptibility matrix}

Formally, the dynamic susceptibility matrix is given by the
Kubo formula
\eq
\chi_{ab}(\omega)
= \int_{-\infty}^{0} \dif t\,\,
\me^{-i\omega t}
\expvalequil{i
\comm{-\phi_{b}(t)}{\phi_{a}(0)}}
\en
however the integral diverges at $t=0$.
The unrenormalized boundary susceptibilities are divergent.
The Fourier transform of the formal Kubo formula is
\eq
\chi_{ab}(\eta) =
\frac1{2\pi} \int_{-\infty}^{\infty} \dif
\omega\,\,
\frac{G_{ab}(\omega)}{\omega-\eta-i\epsilon}
\en
which diverges, in general, since $G_{ab}(\omega)$
can grow as fast as $\omega$ for large $\omega$.
Renormalizing the boundary susceptibilities requires two
subtractions:
a constant subtraction and a linear subtraction,
proportional to $\eta$.
A renormalized dynamic
susceptibility matrix is defined by
\eq
\chiren_{ab}(\eta) =
\frac1{2\pi} \int_{-\infty}^{\infty} \dif
\omega\,\,
\left [
\frac1{\omega-\eta-i\epsilon}
- \frac{\omega+\eta+i\epsilon}{\omega^{2}+\omega_{1}^{2}}
\right ] G_{ab}(\omega)
\:.
\label{eq:chiab}
\en
The subtractions are chosen so that
$\chiren_{ab}(\omega)$ will be
compatible with the natural susceptibilities
$F_{a}(\omega)$ and $F(\omega)$:
\begin{eqnarray*}
F_{a}(\omega) &=& \chi_{ab}(\omega) \beta^{b}\\
F(\omega) &=& \chi_{ab}(\omega) \beta^{a} \beta^{b}
\:.
\end{eqnarray*}
$F_{a}(\omega)$ and $F(\omega)$ are natural in the sense
that they are constructed
without arbitrary subtractions,
in terms of the chiral energy currents outside the
boundary.
$\chiren_{ab}(\omega)$ is a dynamic susceptibility matrix
in the sense that (1) it
is analytic in the upper half-plane,
(2) it satisfies,
on the real axis,
$$
\chiren_{ab}(\omega)-\overline{\chiren_{ba}(\omega)}
= i G_{ab}(\omega)
= \int_{-\infty}^{\infty} \dif t\,\,
\me^{-i\omega t}
\expvalequil{i
\comm{-\phi_{b}(t)}{\phi_{a}(0)}}
$$
and (3) its static limit,
\eq
\chiren_{ab}(0) = T g_{ab}
\:,
\en
is a positive symmetric matrix.
The metric $g_{ab}$ on the space of boundary couplings
is the renormalized static susceptibility matrix
for the boundary, divided by the temperature.

\subsection{The imaginary time formula for the metric}

The imaginary time formula for the metric $g_{ab}$ is \cite{FK}
\eq
T \geucl_{ab} =
\int_{0}^{ \beta} \dif \tau \,\,
\expvalc{\phi_{b}(-i\tau)\,\phi_{a}(0)}
\left [
1- \cos(\omega_{1} \tau )
\right ]
\:.
\en
From equation (\ref{eq:Gabtildephi}),
$$
\expvalc{\phi_{b}(-i\tau)\,\phi_{a}(0)}
=
\frac1{2\pi} \int \dif \omega\,\,
\me^{\omega \tau}
\frac{G_{ab}(\omega)}{\me^{\beta  \omega}-1}
$$
and therefore
\begin{eqnarray*}
T \geucl_{ab} &=&
\frac1{2\pi} \int \dif \omega\,\,
\frac{G_{ab}(\omega)}{\me^{\beta  \omega}-1}
\int_{0}^{ \beta} \dif \tau \,\,
\me^{\omega \tau}
\left [
1- \frac12\me^{i\omega_{1}\tau}- \frac12\me^{-i\omega_{1}\tau}
\right ]
\\
&=&
\frac1{2\pi} \int \dif \omega\,\,
\frac{
G_{ab}(\omega)}{\omega(1+\omega_{1}^{-2}\omega^{2})}
\end{eqnarray*}
which is exactly equation (\ref{eq:gab}) for $T g_{ab}$,
since $G_{ab}(-\omega) = - \overline{G_{ab}(\omega)}$.
So the real time and
imaginary time formulas
for the metric $g_{ab}$ are equivalent.
\section{Estimate on $ds/dT$ using dispersion formula}
\label{sect:estimate}

Combining formula (\ref{dsdt}) with the dispersion relation
(\ref{disprel}) we can write
\begin{equation}\label{dsdt2}
\frac{ds}{dT} = 2\pi\int\limits_{-\infty}^{\infty}d\omega\,
\frac{-{\rm \bf Re} F(\omega, T)}{\omega^{2}(4\pi^{2}T^{2} +
\omega^{2})} \, .
\end{equation}
Here we included the
temperature argument in the notation of the response
function: $F(\omega)=F(\omega, T)$.  By property 2 from
section \ref{propF} the function ${\rm \bf Re} F(\omega, T)$
is even on the real axis.  Since $F(0, T)=0$ (property 3)
the integral on the right hand side of (\ref{dsdt2}) is well
defined for $T>0$.  In the limit $T\to 0$ however the poles
at $\omega=0$ and $\omega=\pm i\omega_{1}$ coalesce.  To
separate the dangerous part we rewrite the right hand side
as
\begin{equation}\label{dsdt3}
\frac{ds}{dT} =
2\pi\int\limits_{-\infty}^{\infty}d\omega\, \frac{-{\rm \bf
Re} F(\omega, T) + \frac{\omega^{2}}{2}{\rm \bf Re} F''(0,
T)}{\omega^{2}(4\pi^{2}T^{2} + \omega^{2})} -\frac{\pi}{2 T}
F''(0, T) \, .
\end{equation}
Here the integral on the right hand side converges as $T\to
0$ to a constant
\begin{equation}
f=-2\pi \int\limits_{-\infty}^{+\infty}d\omega\,
{\rm \bf Re} F(\omega,0) {\cal P}
\left ( \frac{1}{\omega^{4}} \right )
\end{equation}
where ${\cal P}\left ( \frac1{\omega^{4}}\right )$ is the
standard (even) distribution regularizing the function
$1/\omega^{4}$.  The first term in (\ref{dsdt3}) is
therefore integrable at $T=0$.

Let us look now at the second term on the right hand side of
(\ref{dsdt3}).  By comparing formula (\ref{tTEucl}) with the
$T\to 0$ limit of formula (\ref{FRcorr}) we obtain
\begin{equation}\label{spec_fns}
A_{\theta R}(\omega) = -\lim_{T\to 0} F_{R}(\omega ,T) \, .
\end{equation}
It follows then from (\ref{tT''}) that
\begin{equation} \label{eq:Fpp}
\label{F''T0} \lim_{T\to 0} F''(0, T) = 0
\end{equation}
However general analysis stops here as we have no control in
general over how fast $ F''(0, T)$ vanishes as $T\to 0$ and
thus cannot conclude whether $$ \frac{{\rm \bf Re}
F''(0,T)}{T} $$ is integrable in a neighborhood of $T=0$.

Equation (\ref{eq:Fpp}) implies that $T\dif s/\dif T \to 0$
as $T\to 0$.  Note that in deriving this we used in the essential way the consequences of the
bulk conformal invariance on a half plane, that is the condition $\langle 0|Q=0$
discussed in section 4.

Although the above manipulations do not lead to demonstrating  the boundedness of $s$ they boil
down the problem to having an estimate on the zero temperature limit
 of  $F''(0,T)$.  For models possessing an asymptotic power expansion at  small temperature
  (such as e.g. the Ising model with a boundary magnetic field) the existence
 of a lower bound on $s$ follows from (\ref{dsdt3}), (\ref{F''T0}).

\begin{center} {\large \bf Acknowledgments} \end{center}

We are grateful to Gregory Moore for useful discussions. This research was supported in part by
DOE grant DE-FG02-96ER40949.


\appendix
\renewcommand{\theequation}{\Alph{section}.\arabic{equation}}
\setcounter{equation}{0}

\section{$\langle \theta \theta\rangle_{c}$ decays at infinity from bulk conformal invariance at
$T>0$ }
\label{app:A}
From (\ref{Ln}) and (\ref{FRcorr}) we obtain
\begin{eqnarray}
-\frac{2\pi}{\beta}e^{-2\pi n x/ v\beta}\langle 0|L_{n}\theta(0)
|B\rangle &=&
\int\limits_{0}^{\beta}d\tau\, e^{2\pi in\tau/\beta}
\langle T_{R}(x+i\tau)\theta(0)\rangle_{eq} \nonumber\\
&=& \frac{1}{2\pi}\int\limits_{0}^{\beta}d\tau\, e^{-2\pi
in\tau/ \beta} \int\limits_{-\infty}^{+\infty}d\omega\,
e^{\omega(-\tau +ix/v)}
\frac{-iF_{R}(\omega)}{(e^{-\beta\omega}-1)} \nonumber \\
&=&\frac{1}{2\pi}\int\limits_{-\infty}^{+\infty}d\omega\,
e^{i\omega x/v} \frac{iF_{R}(\omega)}{(\omega - 2\pi in/\beta)}
\end{eqnarray}
where on the left hand side $\bra{0}$ is the conformal bulk
vacuum, $|B\rangle$ is the boundary state representing our
boundary condition on a cylinder of circumference $\beta$.
Here we use a representation for correlators corresponding
to quantization with $x$ being a Euclidean time.  Since
$\bra{0}L_{n} =0$ for $n\le 1$, we have
\eq
\int\limits_{-\infty}^{+\infty}d\omega\, e^{i\omega x/v}
\frac{F_{R}(\omega)}{(\omega - 2\pi in/ \beta)}=0 \, , \quad
n\le 1\, , x>0 \, .
\en
Taking an appropriate linear combination of the above equations with $n=-1,0,1$ we get\
\begin{equation} \label{A1}
\int\limits_{-\infty}^{+\infty} d\omega\,
e^{i\omega x/v}
\frac{F_{R}(\omega)}{\omega(\omega^{2} + 4\pi^{2} / \beta^{2})}=0 \, , \quad x>0 \, .
\end{equation}
The limit $x\to 0$ gives a sum rule
\begin{equation}
\int\limits_{-\infty}^{+\infty}d\omega\,
\frac{F_{R}(\omega)}{\omega(\omega^{2} + 4\pi^{2} / \beta^{2})}=0 \, .
\label{sumrule}
\end{equation}
As proved in section 5, the function
${\bf \rm Re}F_{R}(\omega)$ is even on the real line
and the function ${\bf
\rm Im}F_{R}(\omega)$ is odd.  Thus equation (\ref{A1})
implies that
\begin{equation} \label{eqss}
\int\limits_{-\infty}^{+\infty}d\omega\,
\frac{[\cos(\omega x/v){\rm
\bf Im}F_{ R}(\omega) + \sin(\omega x/v){\rm \bf Re}F_{
R}(\omega)]} {\omega(\omega^{2} + 4\pi^{2} /\beta^{2})}=0
\end{equation}
for $x>0$.
The spectral function $F_{R}(\omega)=F_{R}(\omega, T)$ is related to the zero temperature spectral function
as in (\ref{spec_fns})
$$
\lim_{T\to 0}F_{R}(\omega, T) = -A_{\theta R}(\omega) \, .
$$
Thus in the limit $T\to 0$  equations (\ref{eqss}) (assuming the limit commutes with the integration) imply that
the functions ${\bf \rm Im} A_{R\theta}(\omega)/\omega^{3}$
and ${\bf \rm Re} A_{R\theta}(\omega)/\omega^{2}$ are
integrable at $\omega=0$.

We did not use here the analyticity of
$F_{R}(\omega)$ in the upper half plane.  That and the sum rule
(\ref{sumrule}) imply that $F_{R}(i2\pi T)=0$.  This zero
can be considered as the sole manifestation of the conformal
invariance at $T>0$.  Another zero is situated at
$\omega=0$.
This allows us to write a subtracted dispersion relation
\begin{eqnarray}\label{disprelT}
&& \frac{{ \rm \bf Re}F_{ R}(\omega)}{(\omega^{2} +
4\pi^{2}/\beta^{2})} =
\frac{1}{2\pi}\int\limits_{-\infty}^{+\infty}d\eta\,
\frac{{ \rm \bf Im}F_{R}(\omega)}{(\eta^{2} +
4\pi^{2}/\beta^{2})}
{\cal P}\left ( \frac{1}{\eta - \omega}\right ) \, ,
\nonumber \\
&& \frac{{ \rm \bf Im}F_{ R}(\omega)}{(\omega^{2} +
4\pi^{2}/\beta^{2})} =
-\frac{1}{2\pi}\int\limits_{-\infty}^{+\infty}d\eta\,
\frac{{ \rm \bf Re}F_{ R}(\omega)}{(\eta^{2} +
4\pi^{2}/\beta^{2})}
{\cal P}\left ( \frac{1}{\eta - \omega}\right )
\end{eqnarray}
As $T\to 0$ these dispersion relations at least formally (more on this below)
go into the dispersion relations (\ref{disprel0})
and the sum rule (\ref{sumrule}) goes into the sum rule (\ref{sumrule0}). It was demonstrated in section 4 how the last two
imply (\ref{tT''}). Analogous considerations hold for $F_{L}(\omega)$ and imply (\ref{tT'''}).

It should be stressed that in all these manipulations it is
implicitly assumed that taking the limit $T\to 0$ commutes
with integrations in dispersion relations.  Or equivalently
that the dispersion relations for $T>0$ go to those at $T=0$
in a continuous fashion.  It could happen that there are
singularities of $F_{R}(\omega)$ in the lower half plane
that approach $\omega=0$ as $T\to 0$.  The above conclusions
for the behavior of $A_{\theta R}(\omega)$ at zero then
would be incorrect.  The continuity at $T=0$ of the above
equations is essentially equivalent to the condition of the
asymptotic conformal invariance on a half plane $\langle
0|Q=0$ discussed in section 4.

To conclude we see that, assuming the just discussed continuity at $T=0$,
the finite $T$ conformal invariance implies formulas (\ref{tT''}), (\ref{tT'''}) and, as a consequence,
the vanishing of the
$\langle \theta(\tau)\theta(\tau')\rangle_{c}$ correlator in the infrared.

\setcounter{equation}{0}
\section{A lower bound on boundary entropy of Cardy
states${}^{\mbox{\small 3}}$ }
\footnotetext[3]{The contents of this appendix grew
up from discussions of A.~K. with G.~Moore}
\setcounter{equation}{0}
For rational conformal field theories there is a set of local conformal
boundary conditions that preserve the chiral algebra in the most straightforward way -
Cardy boundary states. These boundary states are constructed via the modular S-matrix
$S_{ij}$ as
$$
|i\rangle = \sum_{j} \frac{S_{ij}}{\sqrt{S_{0j}}} |j\rangle\rangle \,
$$
where $|j\rangle \rangle$ are Ishibashi states. (Note that in the proof of the Verlinde
formula one shows at an intermediate step that $S_{0j}>0$ and thus the division makes sense.)
The boundary entropies of the Cardy states are
$$
g_{i} = \langle 0|i\rangle = \frac{S_{i0}}{\sqrt{S_{00}}} \, .
$$
We are going to show that
\begin{equation} \label{RCFTb}
g_{i}\ge g_{0} =\sqrt{S_{00}} \, .
\end{equation}
The $S$-matrix entries $S_{ij}$ can be considered as a collection of common eigenvectors
of the fusion matrices. By the Perron-Frobenius theorem there is a unique a common eigenvector
whose set of eigenvalues consists of a maximal eigenvalue for each fusion matrix. It is uniquely
characterized by the property that all its entries are positive.
Since $S_{0j}>0$ this eigenvector corresponds to the zero weight and the corresponding collection
of maximal eigenvalues is
$$
\gamma_{i}^{max} = \frac{S_{i0}}{S_{00}} \, .
$$
On the other hand, the inequality (\ref{RCFTb}) translates into
$$
S_{i0}\ge S_{00}
$$
that is what we need to prove that the maximal eigenvalues of the fusion matrices are all greater
or equal to one. To this end we note that the dimension of the Friedan-Shenker vector bundle
\cite{FS} over a genus $g$ Riemann surface with $n$ punctures in
representations $i_{1}, \dots, i_{n}$
is given by the formula \cite{MooreSeib}
\begin{equation}
{\rm dim}{\cal H}(\Sigma_{g};(P_{1}, i_{1}), \dots, (P_{n}, i_{n})) =
\sum_{p}\frac{1}{(S_{0p})^{2(g-1)}}\frac{S_{i_1 p}}{S_{0p}}\cdot \dots \cdot
\frac{S_{i_{n} p}}{S_{0p}}\, .
\end{equation}
For $g=1$ each summand is a  product of all eigenvalues  of the $i_{k}$-th
fusion matrix. Since the number
of punctures and the weights $i_{k}$ can be arbitrary and the left hand side of the above
 equation is a natural number we conclude that the maximal eigenvalues have to be greater than
 $1$. This concludes the proof of (\ref{RCFTb}).

\end{document}